\documentclass[11pt]{article}
\usepackage{amsmath}
\usepackage{graphicx}
\usepackage{amssymb}
\usepackage{amsfonts}
\usepackage{latexsym}
\usepackage{psfig}

\def\beq{\begin{eqnarray}}
\def\eeq{\end{eqnarray}}

\def\de{\delta}

\def\la{\lambda}

\def\si{\sigma}

\begin{document}

\title{Trajectories in a space with a spherically symmetric dislocation}
\author{Alcides F. Andrade\footnote{Email: afandrade@fisica.ufjf.br} $\,$ and
Guilherme de Berredo-Peixoto\footnote{Email: 
guilherme@fisica.ufjf.br} 
\\ \\
Departamento de F\'\i sica, ICE, Universidade Federal de Juiz de Fora \\
Campus Universit\' ario - Juiz de Fora, MG Brazil  36036-330}
\date{}
\maketitle

\begin{quotation}
\noindent
{\large{\bf Abstract.}}
We consider a new type of defect in the scope of linear 
elasticity theory, using geometrical methods. This defect 
is produced by a spherically symmetric dislocation, or  
ball dislocation. We derive the induced metric as well as 
the affine connections and curvature tensors. Since
the induced metric is discontinuous, one can expect 
ambiguity coming from these quantities, due to products
between delta functions or its derivatives, plaguing
a description of ball dislocations based on the Geometric 
Theory of Defects. However, exactly as in the previous case 
of cylindric defect, one can obtain some 
well-defined physical predictions of the induced geometry. 
In particular, we explore some properties of test particle 
trajectories around the defect and show that these trajectories
are curved but can not be circular orbits.



\end{quotation}

\section{Introduction}

The topological defects attract great interest due to the 
applications to condensed matter physics (see, e.g., \cite{ChaiLub,defects} 
for an introduction and recent review). Another elegant approach in 
this area is the geometric theory of defects \cite{KatVol92,KatVol99} 
(see also \cite{Katana05} for the introduction), which is formulated 
in terms of the notions originally developed in the theories 
of gravity. In this framework, we can cite basically two kinds of
defects, described in the view of Riemann-Cartan geometry: disclinations 
and dislocations. This means that the curvature and torsion tensors,
respectively, are interpreted as surface densities of Frank and 
Burgers vectors and thus linked to the nonlinear, generally 
inelastic deformations of a solid. Recently, the qualitatively 
new kind of geometric defect has been described in \cite{tube} (see also 
\cite{schroedinger}), corresponding to a tube dislocation (with cylindrical 
symmetry).

Nevertheless, one can find others well known types of dislocations.
In Elasticity Theory, the dislocations and their physical effects are a matter
of great concern, specially the screw dislocation \cite{screw}. In this paper,
we consider a ball dislocation (or sphere dislocation), which is the same type of defect studied in \cite{tube}, translated to spherical symmetry. Is 
is remarkable that such a problem was not investigated yet. For our purposes, 
the approach will be limited to linear elasticity theory (using Riemann-Cartan 
geometry as a tool). The ball dislocation can be understood by the illustration 
in Figure 1.

\begin{quotation}
\begin{figure}[h,t]
\psfig{file=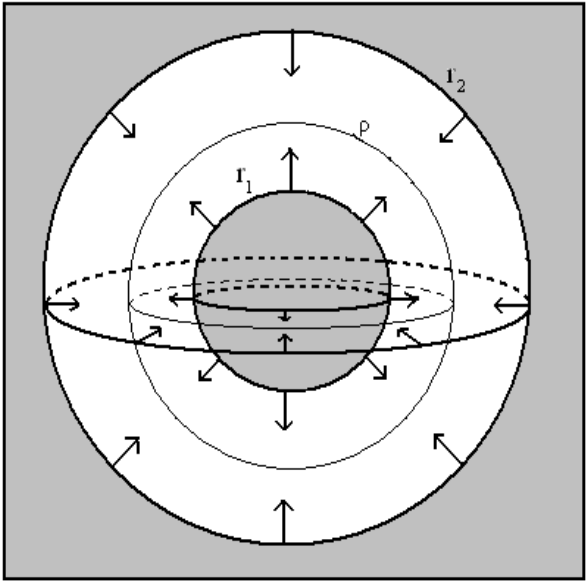,width=8cm,height=8cm}
\centering \caption{Ball dislocation produced by cutting a spherical 
sector, and then indentifying the surfaces $r_1$ and $r_2$ with $\rho$.} 
\end{figure}
\end{quotation}

It is possible to treat this defect as a point defect in the limiting case where
$\rho$ is much smaller than the dimensions of the considered physical system. 
Of course, the ball dislocation provides a more general picture. This paper is 
organized in the following way. In Section 2, we treat mathematically the 
ball dislocation, find the solution of the variable which describes the 
defect, and with the help of Gravity Theory methods, we study the trajectories
of test particles in Section 3. Finally we proceed to our conclusions.

\section{Ball dislocation in linear elasticity theory}

Let us describe the ball dislocations in linear elasticity 
theory. Consider a homogeneous and isotropic 
elastic media as a three-dimensional Euclidean space 
${\mathbb R}^3$ with Cartesian coordinates $\,x^i,\,y^i$, 
where $\,i=1,2,3$. The Euclidean metric is denoted by
$\delta_{ij}={\rm diag}(+++)$. The basic variable in the 
elasticity theory is the displacement vector of a 
point in the elastic media, $u^i(x)$, $\,x\in{\mathbb R}^3$. 
In the absence of external forces, Newton's and Hooke's 
laws reduce to three second order partial differential
equations which describe the equilibrium state of elastic 
media (see, e.g.,~\cite{LanLif70}),
\begin{equation}                                                  
\label{eqsepu}
  (1-2\sigma)\Delta u_i+\partial_i\partial_j u^j=0\,.
\end{equation}
Here $\Delta$ is the Laplace operator and 
the dimensionless Poisson ratio $\si$
\ ($-1\le\si\le 1/2$) \ is defined as 
\begin{equation*}
\si = \frac\la{2(\la+\mu)}\,.
\end{equation*}
The quantities $\la$ and $\mu$ are called the Lame coefficients, which
characterize the elastic properties of media.

Raising and lowering of Latin indices
can be done by using the Euclidean metric, $\delta_{ij}$, 
and its inverse, $\delta^{ij}$.
Eq. (\ref{eqsepu}) together with the corresponding 
boundary conditions enable one to establish the 
solution for the field  $u^i(x)$ in a unique way.

Let us pose the problem for the ball dislocation shown in 
Figure 1. This dislocation can be produced as follows. We cut 
out the thick spherical sector of media located between two concentric 
spherical surfaces of radii $r_1$ and $r_2$ ($r_1<r_2$), move 
symmetrically both cutting surfaces one to the other and 
finally glue them. Due to spherical symmetry of the problem, 
in the equilibrium state the 
gluing surface is also spherical, of radius $\rho$ 
which will be calculated below.

Within the procedure described above and shown in 
Figue 1, we observe the negative ball 
dislocation because part of the media was removed. This 
corresponds to the case of $r_1<r_2$. 
However, the procedure can be applied in the opposite 
way by addition of extra media to ${\mathbb R}^3$. In
this case, we meet a positive ball dislocation and 
the inequality has an opposite sign, $r_1>r_2$.

Let us calculate the radius of the equilibrium 
configuration, $\rho$. This problem is naturally formulated 
and solved in spherical coordinates, $\,r,\,\theta,\,\phi$. Let 
us denote the displacement field components in these
coordinates by $\,u^r,\,u^\theta,\,u^\phi$. In our case, 
$\,u^\theta = u^\phi = 0\,$ due to the symmetry of the problem, 
so that the radial displacement field $u^r(r)$ can be 
simply denoted as $\,u^r(r) = u(r)$.

The boundary conditions for the equilibrium ball
dislocation are
\begin{equation}                                                  
\label{ebotud}
  u\big|_{r=0}=0\,,\qquad 
  u\big|_{r=\infty}=0\,,\qquad 
 \left. \frac {du_{\text{in}}}{dr}\right|_{r=r_*}
 =\left.\frac{du_{\text{ex}}}{dr}\right|_{r=r_*}\,.
\end{equation}
The first two conditions are purely geometrical, and the 
third one means the equality of normal elastic forces 
inside and outside the gluing surface in the equilibrium 
state. The subscripts ``in'' and ``ex'' denote the 
displacement vector field inside and outside the gluing 
surface, respectively.

Let us note that our definition of the displacement vector 
field follows \cite{Katana05}, but differs slightly from 
the one used in many other references. In our notations, 
the point with coordinates $y^i$, after elastic 
deformation, moves to the point with coordinates $x^i$:
\begin{equation}                                                  
\label{eeldef}
y^i\rightarrow x^i(y)=y^i+u^i(x)\,.
\end{equation}
The displacement vector field is the difference between 
new and old coordinates, $u^i(x)=x^i-y^i$. Indeed, we are
considering the components of the displacement vector field, 
$u^i(x)$, as functions of the final state coordinates of 
media points, $x^i$, while in other references they are 
functions of the initial coordinates, $y^i$. The two 
approaches are equivalent in the absence of dislocations 
because both sets of coordinates $x^i$ and $y^i$ cover 
the entire Euclidean space ${\mathbb R}^3$. On the contrary, if 
dislocation is present, the final state coordinates $x^i$ 
cover the whole ${\mathbb R}^3$ while the initial state coordinates
cover only part of the Euclidean space lying outside the 
thick sphere which was removed. For this reason the final 
state coordinates represent the most useful choice here. 

The elasticity equations (\ref{eqsepu}) can be easily 
solved for the case of ball dislocation under consideration. 
Using the Christoffel symbols in order to evaluate the 
expressions for the differential operators (Laplacian and divergence),
equations (\ref{eqsepu}) reduce to the only one non-trivial equation 
\beq
\frac{d^2 u(r)}{dr^2} + \frac{2}{r}\frac{d u(r)}{dr} - 
\frac{2}{r^2} u(r) = 0\,.
\label{eqU}
\eeq
One can remember that only the radial component differs 
from zero. The angular $\theta$ and $\phi$ components of 
equations (\ref{eqsepu}) are identically satisfied. The
general solution for (\ref{eqU}) is given by 
\begin{equation}
  u = \alpha r - \frac{\beta}{r^2}\,,
\label{solu}
\end{equation}
which depends on the two arbitrary constants of integration 
$\alpha$ and $\beta$. Due to the first two boundary conditions 
(\ref{ebotud}), the solutions inside and outside the 
gluing surface are
\beq
\label{elsopt}
\begin{aligned}
  u_{\text{in}} & = \alpha r\,, \quad & \alpha & > 0,
\\
  u_{\text{ex}} & = -\frac{\beta}{r^2}\,, \quad  & \beta & > 0.
\end{aligned}
\eeq
The signs of the integration constants correspond to the 
negative ball dislocation. 
For positive ball dislocation, both integration constants have 
opposite signs. 

Using the solution (\ref{solu}) and the third boundary 
condition (\ref{ebotud}), one can determine the radius 
of the gluing surface,
\begin{equation}                                                  
\label{eglsur}
  \rho = \frac{2 r_2 + r_1}{3}.
\end{equation}
One can see that $\rho$ is not the mean between $r_1$ and $r_2$,
as it is for the cylindrical symmetry defect \cite{tube}. On
the contrary, the gluing surface is located closer to the
external radius $r_2$. After simple algebra, the integration 
constants can be expressed in terms of $\rho$ and the thickness 
of the sphere $l = r_2 - r_1$:
\begin{equation}                                                  
\label{eintco}
\alpha = \frac{2l}{3\rho}
\qquad
\beta = \frac{\rho^2 l}{3}\,.
\end{equation}
It is straightforward also to get
$$
r_1 = \frac{3\rho - 2l}{3}\;\; {\rm and} \;\; 
r_2 = \frac{3\rho + l}{3}\,. 
$$
Observe that as $r_1$ is positive, we must have 
always $\rho > 2l/3$. 

Finally, within the linear elasticity theory, 
eq. (\ref{elsopt}) with the integration constants 
(\ref{eintco}) yields a complete solution for the ball 
dislocation in linear elasticity theory, valid for small 
relative displacements, when \ $l/r_1\ll 1$ and $l/r_2\ll 1$.
It is remarkable that the solution obtained in the 
framework of linear elasticity theory does not depend 
on the Poisson ratio of the media. In this sense, the 
ball dislocation is a purely geometric defect which 
does not feel the elastic properties.

In order to use the geometric approach, we 
compute the geometric quantities of the manifold 
corresponding to the ball dislocation. From the geometric 
point of view, the elastic deformation (\ref{eeldef}) is 
a diffeomorphism between the given domains in the Euclidean 
space. The original elastic media ${\mathbb R}^3$, before the 
dislocation is made, is described by Cartesian coordinates 
$y^i$ with the Euclidean metric $\delta_{ij}$. An inverse
diffeomorphic transformation $x\rightarrow y$ induces 
a nontrivial metric on ${\mathbb R}^3$, corresponding to the ball 
dislocation. In Cartesian coordinates, this metric has the form
\begin{equation}
g_{ij}(x)\,=\,
\frac{\partial y^k}{\partial x^i}\,\frac{\partial y^l}{\partial x^j}
\,\delta_{kl}.
\end{equation}
We use curvilinear spherical coordinates for the ball 
dislocation and therefore it is useful to modify our 
notations. The indices in curvilinear coordinates
in the Euclidean space ${\mathbb R}^3$ will be denoted by 
Greek letters $\,x^\mu$, $\mu=1,2,3$. Then the
``induced'' metric for the ball dislocation in 
spherical coordinates is
\begin{equation}                                                  
\label{eincym}
g_{\mu\nu}(x)\,=\,\frac{\partial y^\rho}{\partial x^\mu}\,
\frac{\partial y^\si}{\partial x^\nu}\,\,\overset{\circ}g_{\rho\si}\,,
\end{equation}
where $\overset{\circ}g_{\rho\si}$ is the Euclidean 
metric written in spherical coordinates. We denote 
spherical coordinates of a point before the dislocation
is made by $\lbrace y, \theta, \phi \rbrace$, where $y$ without 
index stands for the radial coordinate and we take into 
account that the coordinates $\theta$ and $\phi$ do not 
change. Then the diffeomorphism is described by a 
single function relating old and new radial coordinates 
of a point, $y=r-u(r)$, where
\beq
u(r) & = & \frac{2l}{3\rho} r\,,\quad  r < \rho \\
u(r) & = & -\frac{\rho^2 l}{3 r^2}\,,\quad  r > \rho\,. 
\label{edifun}
\eeq
It is easy to see that this function has a discontinuity
$u_{\rm ext} - u_{\rm int} |_{r=\rho} = l$
at the point of the cut. Therefore a special 
care must be taken in calculating the components of induced 
metric. 

It is useful to express $u(r)$ in a way simultaneously valid in both 
domains, $r < \rho$ and $r > \rho$. We have then

\beq
u(r) = \frac{2l}{3\rho} r H(\rho - r) - \frac{\rho^2 l}{3r^2} H(r - \rho)\,,
\eeq
where $H(r-\rho)$ is the Heaviside step function. As $H^\prime(r-\rho) = 
\delta(r-\rho)$, one achieves
\beq
u^\prime(r) = \frac{du(r)}{dr} = v(r) - l \delta(r-\rho)\,,
\eeq
where 
$$
v(r) =  \frac{2l}{3\rho} H(\rho - r) + \frac{2\rho^2 l}{3r^3} H(r - \rho)\,.
$$

By direct calculation of induced metric, by (\ref{eincym}), one can
write the corresponding line element as
\beq
ds^2 = (1 - u^\prime)^2dr^2 + (r - u)^2(d\theta^2 + \sin^2 \theta d\phi^2)\,.
\label{linelement}
\eeq
It is clear that the above expression, besides discontinuous, contains also
a $\de$-function thanks to $u^\prime$. In order to avoid further conceptual
consequences coming from a $\de$-function in the line element\footnote{With 
a $\de$-function in the metric, the Burgers vector can not be defined 
properly as a surface integral \cite{Katana05}.}, we shall drop it, and 
adopt $u^\prime(r) \to v(r)$. In other words, let us consider the line
element
\beq
ds^2_{{\rm int}} & = & \left(1 - \frac{2l}{3\rho}\right)^2 
(dr^2 + r^2d\theta^2 + r^2 \sin^2\theta d\phi^2 )\,, \nonumber \\
ds^2_{{\rm ext}} & = & \left(1 - \frac{2\rho^2 l}{3 r^3}\right)^2 dr^2 +
\left(1 + \frac{\rho^2 l}{3 r^3}\right)^2
(r^2d\theta^2 + r^2 \sin^2\theta d\phi^2 )\,. \nonumber
\eeq    

Notice that the interior space is conformally flat, with a constant 
scale factor, while the exterior metric is not so. Both metrics
are flat (as follows by direct calculation of Riemann tensor), as
they should be (because they were obtained by coordinate transformations
starting from the Euclidean metric). Nevertheless, the whole space is
non-trivial since curvature is non-trivial exactly in the gluing surface.
Next, we are going to investigate the consequences for trajectories of
test particles around the defect.

\section{Trajectories of test particles around the defect}

What are the trajectories of test particles in a space with such
a defect? Of course, the trajectories without defect would be 
straight lines, so we expect deviation from straight lines in
the actual path. How these trajectories can be described? Are there
any possible closed path around the defect? In order to answer these 
questions, let us consider the geodesic equations for both metrics, 
in the interior and in the exterior of gluing surface. The geodesic 
equations read
\beq
\frac{dU^\mu}{d\tau} + \Gamma^\mu\mbox{}_{\rho\la} U^\rho U^\la = 0\,,
\eeq
where $U^\mu = dx^\mu / d\tau = (1,\, \dot{r},\, \dot{\theta},\, \dot{\phi})$
($c = 1$ and dot means derivative with respect to $\tau$) and 
$$
\Gamma^\mu\mbox{}_{\rho\la} = \frac12 g^{\mu\si} 
\left(\partial_\la g_{\si\rho} + \partial_\rho g_{\si\la} -
\partial_\si g_{\rho\la}\right)\,.
$$

Let us remmember that if only dislocations are present, then only
torsion (without Riemannian curvature) is found in the geometric
approach -- only the Burgers vector is non-trivial. But for
practical purposes, one can treat the problem in the reverse
way, considering only Riemannian curvature, because both 
approaches are equivalent. This equivalence is very well-known in
telleparalelism (see, e.g., \cite{pereira}).

Inside the defect, the geodesics are straight lines and thus we
shall consider only the exterior metric. By direct calculation, 
the geodesic equations, outside the gluing surface, can be written 
as
\beq
\ddot{r} + \frac{6\rho^2 l}{r(3r^3 - 2\rho^2 l)}\dot{r}^2 -
\frac{r(\rho^2 l + 3r^3)}{3r^3 - 2\rho^2 l}
\left(\dot{\theta}^2 + \sin^2\theta \dot{\phi}^2\right) = 0\,,
\label{geod1}
\eeq
\beq
\ddot{\theta} - \frac{4\rho^2 l - 6r^3}{r(\rho^2 l + 3r^3)}\dot{r}\dot{\theta}
- \sin\theta\cos\theta\dot{\phi}^2 = 0\,,
\label{geod2}
\eeq
\beq
\ddot{\phi} - \frac{4\rho^2 l - 6r^3}{r(\rho^2 l + 3r^3)}\dot{r}\dot{\phi} 
+ \frac{2\cos\theta}{\sin\theta}\dot{\phi}\dot{\theta} = 0\,.
\label{geod3}
\eeq
The denominator appearing on (\ref{geod1}) is always positive,
because from condition $\rho > 2l/3$, one gets $3r^3 - 2\rho^2 l
> 3r^3 - 3\rho^3 > 0$.

An interesting feature that we can understand is that, if $\dot{r} = 0$,
then the test particle should be necessarily at rest. This follows from
equation (\ref{geod1}). This means that if the test particle is moving, so 
its radial coordinate must be changing: there is no possible trajectory
confined in a spherical surface. In other words, one can say that such
a geometrical defect can not serve as an alternative description of 
gravitating objects (around which we know there are permitted circular 
orbits). However, other kinds of effects, in condensed matter physics,
for example, can not be ruled out.

A test particle can follow also a radial path, defined as any trajectory described by $\dot{\theta} = \dot{\phi} = 0$. To see that,
let us consider $\dot{\theta} = \dot{\phi} = 0$, such that the radial 
equation (\ref{geod1}) reads
$$
\ddot{r} = - \frac{6\rho^2 l}{r(3r^3 - 2\rho^2 l)}\dot{r}^2\,,
$$
which can be solved as
\beq
\dot{r} = \frac{Kr^3}{3r^3 - 2\rho^2 l}\,,
\label{soldotr}
\eeq 
where $K$ is an arbitrary integration constant. This trajectory is 
a straight line, and the particle's radial velocity is such that it
has greater absolute values near the defect. This effect is illustrated
in Figure 2, where we plot the coordinate $r(t)$ against $t$ (the coordinate system is centered at the defect), based on the integration of (\ref{soldotr}),
$$
r + \frac{\rho^2 l}{3r^2} + a\tau + b = 0\,,
$$ 
where $a$ and $b$ are integration constants. We see in Figure 2
the effect of decreasing the absolute velocity as particle gets
away from the defect (if there was no defect, the curve would be 
a straight line). Notice that as much 
the particle is away from the defect, its velocity approaches
a constant value, as one should expect (for $r > > \rho$, 
kinematics is the same from a space without defect). 

\begin{quotation}
\begin{figure}[h,t]
\psfig{file=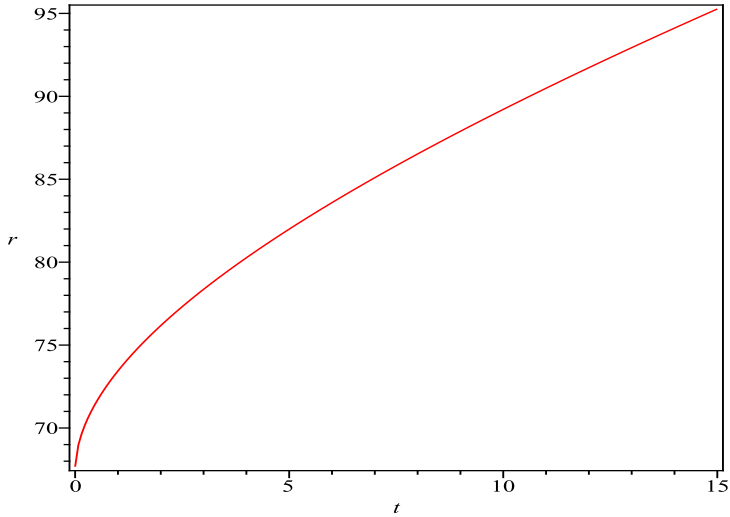,width=10cm,height=7cm}
\centering \caption{Radial curve for $\rho = \frac{203}{3}$ and $\dot{r}_{0} = 50$.} 
\end{figure}
\end{quotation}

To conclude, let us consider $\theta = \pi/2$ ($\dot{\theta} = 0$)
and $\dot{\phi} \neq 0$. The geodesic equations read
\beq
\ddot{r} + \frac{6\rho^2 l}{r(3r^3 - 2\rho^2 l)}\dot{r}^2 -
\frac{r(\rho^2 l + 3r^3)}{3r^3 - 2\rho^2 l}\dot{\phi}^2 = 0\,,
\label{geod4}
\eeq
\beq
\ddot{\phi} - \frac{4\rho^2 l - 6r^3}{r(\rho^2 l + 3r^3)}
\dot{r}\dot{\phi} = 0\,.
\label{geod5}
\eeq
The last equation can be integrated and we obtain
\beq
\dot{\phi} = \frac{A r^4}{(3r^3 + \rho^2 l)^2}\,,
\label{dotphi}
\eeq
where $A$ is some integration constant. Far away from the
defect, we see that $\dot{\phi}$ is proportional to $r^{-2}$.

Substituting (\ref{dotphi}) into (\ref{geod4}), we have
$$
\ddot{r} + \frac{6\rho^2 l}{r(3r^3 - 2\rho^2 l)}\dot{r}^2 -
\frac{r(\rho^2 l + 3r^3)}{3r^3 - 2\rho^2 l}
\frac{A^2 r^8}{(3r^3 + \rho^2 l)^4} = 0\,.
$$

The above equation can be numerically integrated, and
we can learn that the behavior of radial velocity is very
similar to the one described by (\ref{soldotr}), which
can be seen in Figure 3 (both drawn with the help of
MAPLE software).

\begin{quotation}
\begin{figure}[h,t]
\psfig{file=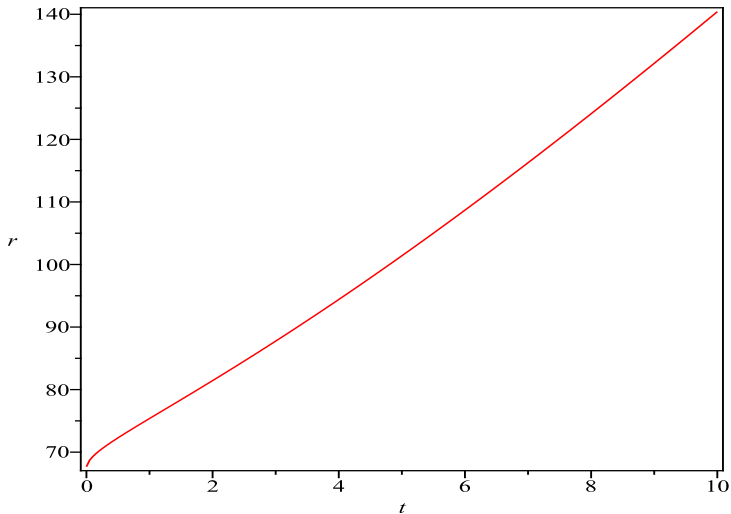,width=10cm,height=7cm}
\centering \caption{Radial curve for $\rho = \frac{203}{3}$ and $\dot{r}_{0} = 50$.} 
\end{figure}
\end{quotation}

One can extract an interesting information from (\ref{dotphi}),
together with the natural assumption that the absolute velocity
decreases in time (as suggested by (\ref{soldotr})).
As the solution $\dot{\phi} = (A/9)r^{-2}$ corresponds to straight
lines (geodesics in flat space without defect), the actual 
derivative $\dot{\phi}$ given by (\ref{dotphi}) decays more slowly 
comparing to the straight line case. This means that the actual
path must deviate from a straight line, curving to the side of 
the defect. In other words, the path of a test particle is
deflected around a defect in a similar way of the gravitational 
deflection.

\section*{Conclusions}

We study a new kind of defect, which we call ball dislocation, using 
geometrical methods in linear elasticity theory. Whenever the displacement
vector (whose discontinuity characterizes the defect) is small comparing
to natural dimensions of some physical system, the linear elasticity theory
is suitable, and the formalism of Geometric Theory of Defects can be disconsidered. Moreover, we consider a single defect and no other complicated
configurations, as a continuous distribution of defects (for which the
Geometric Theory of Defects is required and well-suited).

Nevertheless, it is interesting to investigate the formulation of Geometric Theory of Defects for our problem. In doing so, we find that a direct (naive) 
application of this formulation are faced to ambiguity problems, in contrast to 
other kinds of defects (see \cite{tube}). The corresponding calculations are 
given in the Appendix.

Some interesting properties can be seen in the trajectories of free 
classical particles which follow the geodesic equations in the presence
of spherical defect. Among the properties of such 
motion, we show that any orbit (around the defect) confined on a sphere is forbidden. The circular orbit is a particular case. In the same time we
know that circular orbits are permitted in gravitating systems; thus, according
to at least this feature, the kinematical effects of a defect should not be completely identified with gravitational effects. On the other hand, all 
trajectories are deflected near the defect, in an analogous way of 
gravitating systems.

One can ask if such a defect could describe some real condensed matter
system, where other effects than gravity are dominating. In this case,
we have a geometric description which mimics condensed matter effects 
from electrodynamics. This question is open, but the present article
is a first step in studying the issue. It would be natural to identify
each atom with a defect, and the effects on quantum particles (e.g., 
Dirac fermions) will be an interesting problem addressed to future
works.     

\subsection*{ACKNOWLEDGEMENTS}

AFA acknowledges the CAPES for the schoolarship support and
GBP thanks FAPEMIG and CNPq for financial support. We would like to express 
our gratitude to Ilya Shapiro (UFJF) for useful discussions.

\section*{Appendix}

In order to consider the Geometric Theory of Defects, 
one should start from the induced metric, as derived in linear elasticity 
theory, calculate the corresponding curvature tensors, and identify 
the Einstein tensor with the energy-momentum tensor in the geometric
dynamical equations (which is the Einstein equations) \cite{Katana05}
(see also \cite{tube}).

On can write the line element (\ref{linelement}) in the form
\beq
ds^2 = (1 - v)^2dr^2 + (r - u)^2(d\theta^2 + \sin^2 \theta d\phi^2)\,.
\label{A1}
\eeq
and calculate the components of the curvature tensor, given by
\beq
R_{\mu\nu\rho\si} = \partial_{\mu}\Gamma_{\nu\rho\si} - \Gamma_{\nu\rho}\mbox{}^\la 
\Gamma_{\mu\si\la} - (\mu \leftrightarrow \nu).
\eeq
so
\begin{eqnarray}
R_{r \theta r \theta} &=& -l(r - u)\left[ \delta '(r - \rho) + 
\frac{v'}{(1 - v)}\delta(r - \rho)\right] ,\\
R_{r \phi r \phi} &=& -l(r - u)\left[ \delta '(r - \rho) + 
\frac{v'}{(1 - v)}\delta(r - \rho)\right]\sin ^{2}\theta ,\\
R_{\theta \phi \theta \phi} &=& -l(r - u)^{2}\left[ \frac{2}{(1 - v)}\delta(r - \rho) + 
\frac{l}{(1 - v)^{2}}\delta ^{2}(r - \rho)\right]\sin ^{2}\theta ,
\end{eqnarray}
with $\delta '(r - \rho) = d \delta(r - \rho)/dr$. 
The components of the Ricci tensor are given by:
\begin{equation}
R_{\nu \beta} = R^{\mu}{}_{\nu \mu \beta} . 
\end{equation}
Hence, for the line element (\ref{A1}):
\begin{eqnarray}
R_{r r} &=& -\frac{2l}{(r - u)}\left[ \delta '(r - \rho) + 
\frac{v'}{(1 - v)}\delta(r - \rho)\right] ,\\
R_{\theta \theta} &=& -\frac{l}{(1 - v)^{3}} \{(1 - v)(r - u)\delta '(r - \rho) + 
\nonumber \\ & & + \left[ v'(r - u) + 2(1 - v)^{2}\right]  \delta(r - \rho) + 
\nonumber \\ & & + l(1 - v)\delta ^{2}(r - \rho)\} ,\\
R_{\phi \phi} &=& -\frac{l}{(1 - v)^{3}}\{  (1 - v)(r - u)\delta '(r - \rho) + 
\nonumber  \\ & & + \left[ v'(r - u) + 2(1 - v)^{2}\right] \delta(r - \rho) + 
\nonumber  \\ & & + l(1 - v)\delta ^{2}(r - \rho)\} \sin ^{2}\theta .
\end{eqnarray}
The scalar curvature $R$ is given by:
\begin{equation}
R = R^{\mu}{}_{\mu}. 
\end{equation}
Hence
\begin{eqnarray}
R &=& - \frac{4l}{(r - u)^{2}(1 - v)^{3}} \{ (r - u)(1 - v)\delta '(r - \rho) + 
\nonumber \\ & & + \left[ v'(r - u) + (1 - v)^{2}\right]\delta(r - \rho) \} - 
\nonumber  \\ & & - \frac{2l^{2}}{(r - u)^{2}(1 - v)^{2}}\delta ^{2}(r - \rho). 
\end{eqnarray}

Notice that the curvature is non-trivial only in the gluing surface. 
Moreover, these quantities are also ambiguous because of the appearance of the 
product of $\delta$-function for discontinuous functions, and the ambiguity is
not cancelled in the calculation of Einstein tensor.

\end{document}